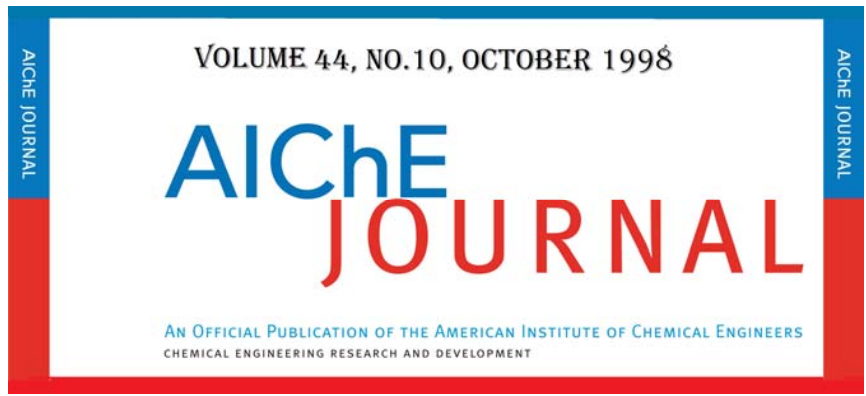

# Surface Tension Prediction for Liquid Mixtures

**Joel Escobedo and G.Ali Mansoori**[(*)]
University of Illinois at Chicago
(M/C 063) Chicago, IL 60607-7052, USA

*key words: Surface tension, organic fluids, mixtures, prediction, new mixing rules*

## Abstract

Recently Escobedo and Mansoori (*AIChE J.* **42**(5): 1425, 1996) proposed a new expression, originally derived from statistical mechanics, for the surface tension prediction of pure fluids. In this report, the same theory is extended to the case of mixtures of organic liquids with the following expression,

$$\sigma_m = \left[ \left( 1-T_{r,m} \right)^{0.37} T_{r,m} \exp \left( 0.30066/T_{r,m} + 0.86442\, T_{r,m}{}^9 \right) \left( \mathscr{P}^L{}_{0,m}\, \rho_{L,m} - \mathscr{P}^V{}_{0,m}\, \rho_{v,m} \right) \right]^4$$

$$\mathscr{P}_{0,m} = \left\{ \sum_i \sum_j \left[ x_i x_j \left( P_{c_{ij}} / T_{c_{ij}} \right)^{7/3} \mathscr{P}^4{}_{0,ij} \right] \right\}^{1/4} \left[ \sum_i \sum_j \left( x_i x_j\, T_{c_{ij}} / P_{c_{ij}} \right) \right]^{7/12}$$

$$\mathscr{P}_{0,ij} = \left( 1-m_{ij} \right) \left( \mathscr{P}_{0,i} \mathscr{P}_{0,j} \right)^{1/2}$$

where $\sigma_m$ is the surface tension of the mixture. $T_{r,m}=T/T_{c,m}$ and $T_{c,m}$ is the pseudo-critical temperature of the liquid mixture. $\rho_{L,m}$ and $\rho_{v,m}$ are the equilibrium densities of liquid and vapor, respectively. $\mathscr{P}^L{}_{0,m}$ and $\mathscr{P}^V{}_{0,m}$ are the temperature-independent compound-dependent parameters for the liquid and vapor, respectively. $P_{c,i}$ , $T_{c,i}$ and $x_i$ are the critical pressure, critical temperature and the mole fraction for component **i**, respectively. $\mathscr{P}_{0,i}$ is the temperature-independent parameter for component **i** and $m_{ij}$ is the unlike interaction parameter.

Using the proposed equation surface tensions of 55 binary mixtures are predicted within an overall 0.50 AAD% which is better than the other available prediction methods. When the $\mathscr{P}_{0,i}$'s are made compound-insensitive using a corresponding states expression the surface tension of all the 55 binary mixtures are predicted within an overall 2.10 AAD%. The proposed model is also applicable to multicomponent liquid mixtures.

---

(*) Corresponding author, email: *Mansoori@uic.edu*

## Introduction

The surface tension of a liquid mixture is not a simple function of the surface tensions of the pure liquids. Also, the composition of the bulk phase and the composition at the vapor-liquid interface are not always the same. At the interface, there is migration of the species having the lowest surface tension, or free energy per unit area, at the temperature of the system. This migration at the interface results in a liquid-phase rich in the component with the highest surface tension and a vapor phase rich in the component with the lowest surface tension (Teixeira *et al.*, 1992, Croxton, 1986, Carey *et al*, 1980; Winterfield *et al.*, 1978; Massoudi and King, 1975, Wertheim, 1976). According to Chapela *et al.*, (1977) the more volatile component is adsorbed from the mixture and extends about one atomic diameter into the gas phase beyond the other species.

In most instances the bulk composition is known. However, the composition at the vapor-liquid interface is unknown. Therefore, there is a need for relating the surface tension of mixtures to bulk-phase concentrations and properties (*i.e.* densities).

In general, there have been several approaches for estimating the surface tension of mixtures (i) based on empirical or semi-empirical thermodynamic relations suggested first for pure fluids and (ii) based on statistical mechanical grounds.

In a recent publication Escobedo and Mansoori (1996) presented a set of new expressions for predicting the surface tension of pure fluids. These new expressions were developed from the statistical mechanical basis of the Macleod equation. It was shown that their new expressions represent rather accurately the surface tension of a variety of pure fluids over a wide range of temperatures. In this paper, the expressions proposed originally by Escobedo and Mansoori (1996) are extended to the case of mixtures.

## Theory

For the surface tension of pure organic fluids, $\sigma$, Escobedo and Mansoori (1996), proposed,

$$\sigma = \left[ \mathscr{P}(\rho_l - \rho_v) \right]^4 \tag{1}$$

$$\mathscr{P} = \mathscr{P}_0 \left( 1 - T_r \right)^{0.37} T_r \, \text{Exp} \left( 0.30066/T_r + 0.86442 \, T_r^9 \right) \tag{2}$$

where $\mathscr{P}$ is known as parachor, $\rho_l$ is the liquid density, $\rho_v$ is the vapor density, $T_r$ is the reduced temperature and $\mathscr{P}_0$ is a constant which is a function of the characteristics of the molecule under consideration. This new expression, derived from statistical mechanics, represents the experimental surface tension of 94 different organic compounds within 1.05 AAD%. In addition,

$$\mathscr{P}_0 = 39.6431 \left[ 0.22217 - 2.91042 \times 10^{-3} \left( \mathscr{R}_c^* / T_{br}^2 \right) \right] T_c^{13/12} / P_c^{5/6} \tag{3}$$

was proposed as a corresponding-states expression for the temperature-independent,

compound-dependent parameter $\mathcal{P}_0$. In this equation, $\mathcal{R}^* = \mathcal{R}_m/\mathcal{R}_{m,ref}$; $\mathcal{R}_m$ is the molar refraction given by $\mathcal{R}_m = (4\pi N_A/3)[\alpha + \mu^2 f(T)] = (1/\rho)(n^2-1)/(n^2+2)$; $\alpha$ and $\mu$ are the polarizability and the dipole moment of the molecule; $N_A$ is the Avogadro's number, n is the sodium-D light refractive index of a liquid at 20¡ C; $\mathcal{R}_{m,ref}$=6.987 is the molar refraction of the reference fluid (methane). When this generalized expression is used, surface tensions for all the 94 pure compounds can be predicted within an overall 2.57 AAD% at all temperatures investigated.

Exact statistical mechanical extension of equations 1-3 to mixtures results in quite complex expressions. However, using the principles of the conformal solution theory (Massih and Mansoori, 1983) the following generalization of Eq. 2 for the case of mixtures will be concluded,

$$\sigma_m = \left[ \left(1 - T_{rm}\right)^{0.37} T_{r,m} \exp\left(0.30066/T_{r,m} + 0.86442\, T_{r,m}^{\,9}\right) \left(\mathcal{P}^L_{0,m}\, \rho_{L,m} - \mathcal{P}^V_{0,m}\, \rho_{v,m}\right) \right]^4 \qquad (4)$$

where $\sigma_m$ is the surface tension of the mixture; $\mathcal{P}^L_{0,m}$ and $\mathcal{P}^V_{0,m}$ are the temperature-independent conformal parameters for the liquid and vapor phase, respectively; $T_{r,m}$ is the pseudo-reduced temperature for the mixture (*i.e.* $T_{r,m} = T / T_{c,m}$); $T_{c,m}$ is the pseudo-critical temperature of the liquid mixture); and $\rho_{L,m}$ and $\rho_{v,m}$ are the equilibrium densities of the liquid and vapor phase, respectively. The aim of this report is the application of Eq. 4 for the prediction of the surface tension of mixtures, over the whole concentration range from the temperature-independent parameter $\mathcal{P}_0$ for the pure components. In order to apply Eq. 4 to mixtures we need the following,

1. Accurate prediction of phase equilibrium properties (*i.e.* composition of the liquid and vapor phases, vapor and liquid densities, and saturation pressures).
2. Mixing rules for the temperature-independent parameter $\mathcal{P}_0$ in order to calculate the temperature-independent parameters $\mathcal{P}^L_{0,m}$ and $\mathcal{P}^V_{0,m}$ for the liquid-and vapor-phase mixtures.

**Prediction of Phase Equilibrium Properties**

In this work, the equilibrium properties for a multicomponent mixture were calculated in two steps.

First, the Peng-Robinson (PR) equation of state was used to predict the vapor pressure and liquid- and vapor-phase compositions. The PR equation of state was selected for this purpose because its parameters are known for many pure compounds and because its interaction parameters are available for numerous binary mixtures. Since the constants of the PR equation are fitted to vapor pressure data it is reliable for vapor-liquid-equilibrium (VLE) calculations. Second, the predicted compositions and vapor pressure by the PR equation were in turn used with the Riazi-Mansoori (RM) equation of state (described below) to predict the densities of the liquid and vapor phases. The RM equation of state yields more accurate density predictions than the PR equation.

The calculations described above have been incorporated into a comprehensive computer package for vapor-liquid-equilibrium (VLE) calculations. This computer package has been used to perform the routine bubble-point and flash calculations.

The PR equation of state is given by,

$$P = RT / (V - b) - a\alpha /(V^2 + 2bV - b^2) \qquad (5)$$

where P, T, and $V=1/\rho$ are pressure, temperature, and molar volume, respectively; R is the universal gas constant; $a = 0.45724\ R^2 T_c^2 / P_c$; $b = 0.07780\ RT_c / P_c$; $T_c$ is the critical temperature; $P_c$ is the critical pressure; $\alpha = [1 + (0.37464 + 1.5422\ \omega - 0.26992\ \omega^2)(1 - T_r^{0.5})]^2$; $\omega$ is the acentric factor; $T_r = T/T_c$ is the reduced temperature. The mixing rules for the temperature-dependent term (**a** $\alpha$) and constant **b** are given by: $(a\ \alpha)_m = \Sigma\Sigma\ [x_i x_j (a\ \alpha)_{ij}]$ and $b_m = \Sigma\ (x_i\ b_i)$, respectively; $(a\ \alpha)_m$ is the temperature-dependent parameter for the mixture; $b_m$ is the volume-related constant for the mixture; and $x_i$ is the mole fraction of species **i** in the mixture. The cross-parameters are calculated using the following expression, $(a\ \alpha)_{ij} = (1-k_{ij})[(a\alpha)_i (a\alpha)_j]^{1/2}$; where $k_{ij}$ is a binary interaction parameter. This interaction parameter is calculated using a correlation recently proposed by Gao, 1992: $(1 - k_{ij}^{PR}) = [2\ (T_{c_i}\ T_{c_j})^{1/2} / (T_{c_i} + T_{c_j})]^{Z_{c_{ij}}}$; $Z_{c_{ij}} = [(Z_{c_i} + Z_{c_j})/2]$; where $Z_{c_i}$ is the critical compressibility factor for pure component **i**.

Bubble-point and flash calculations are performed using the above equation of state following the standard algorithms reported in textbooks (Walas, 1985). The predicted saturation pressure and vapor- and liquid-phase compositions are then used along with the Riazi-Mansoori (1993) equation of state (RM EOS) to calculate the equilibrium densities for the vapor and liquid phases. The RM EOS is given below,

$$P = \rho RT / (1-b\rho) - a\rho^2 / [T^{1/2}\cdot(1+b\rho)] \qquad (6)$$

$a = 0.42748\ R^2 T_c^{2.5}/P_c$     and    $b = (0.08664\ RT_c/P_c)\cdot\delta(\mathcal{R}_c^*, T_r)$

$\delta^{-1} = 1 + \{\ 0.02[1 - 0.92\ \exp(-1000\ |\ T_r - 1\ |)] - 0.035(T_r - 1)\}\ (\mathcal{R}_c^* - 1)$

$\mathcal{R}_c^* = \mathcal{R}_m / \mathcal{R}_{m,ref}$;   $\mathcal{R}_m = (4\pi N_A/3)[\alpha + \mu^2 f(T)] = (1/\rho)(n^2-1)/(n^2+2)$; $\mathcal{R}_{m,ref} = 6.987$

where $\rho$ is the molar density; T, P, $T_c$, $P_c$, $T_r$, and R are as previously defined; $\mathcal{R}_m$ is the molar refraction; $\mathcal{R}_{m,ref}$ is the molar refraction of the reference compound(methane); $\alpha$ and $\mu$ are the polarizability and the dipole moment of the molecule; $N_A$ is the Avogadro's number, n is the sodium-D light refractive index of a liquid at 20¡ C (Riazi and Mansoori, 1993).

In order to apply Eq. 6 to the case of mixtures the van der Waals mixing rules and the molecular expressions for the cross-parameters (Riazi and Mansoori, 1993) have been used,

$$T_{c_m} = \Sigma_i \Sigma_j\ [x_i x_j\ (T_{c_{ij}}^2 / P_{c_{ij}})] / \Sigma_i \Sigma_j\ [x_i x_j\ (T_{c_{ij}} / P_{c_{ij}})] \qquad (6a)$$

$$T_{c_{ij}} = (1 - k_{ij}^{RM}) (T_{c_i} T_{c_j})^{1/2} \tag{6b}$$

$$P_{c_m} = \sum_i \sum_j [x_i x_j (T_{c_{ij}}^2 / P_{c_{ij}})] / \left\{ \sum_i \sum_j [x_i x_j (T_{c_{ij}} / P_{c_{ij}})] \right\}^2 \tag{6c}$$

$$P_{c_{ij}} = 8 T_{c_{ij}} / [(T_{c_{ii}} / P_{c_{ii}})^{1/3} + (T_{c_{jj}} / P_{c_{jj}})^{1/3}]^3 \tag{6d}$$

$$\mathscr{R}^*_m = \sum_i \sum_j (x_i x_j \mathscr{R}^*_{ij}) \tag{6e}$$

$$\mathscr{R}^*_{ij} = [\mathscr{R}^{*1/3}_{ii} + \mathscr{R}^{*1/3}_{jj}]^3 / 8 \tag{6f}$$

In these equations, $T_{c_m}$ is the pseudo-critical temperature of the mixture; $T_{c_i}$ is the critical temperature of pure component **i** in the mixture; $P_{c_m}$ is the pseudo-critical pressure of the mixture; $P_{c_i}$ is the critical pressure of pure component **i** in the mixture; $P_{c_{ij}}$ is a cross-parameter for pressure; $k_{ij}^{RM}$ is the binary interaction parameter (RM stands for Riazi-Mansoori to differentiate it from the interaction parameter used in the PR equation of state); $\mathscr{R}^*_m$ is the dimensionless molar refraction of the mixture; $\mathscr{R}^*_i$ is the dimensionless molar refraction of pure component **i** in the mixture; and $x_i$ is the mole fraction of component **i** in the mixture. The binary interaction parameter, $k_{ij}^{RM}$, is calculated using the following expression,

$$(1 - k_{ij}^{RM}) = (1 - k_{ij}^{PR}) \left\{ 8 [(T_{c_{ii}} / P_{c_{ii}})(T_{c_{jj}} / P_{c_{jj}})]^{1/2} / [(T_{c_{ii}} / P_{c_{ii}})^{1/3} + (T_{c_{jj}} / P_{c_{jj}})^{1/3}]^3 \right\} \tag{6g}$$

**Mixing Rules for the Temperature-independent parameter $\mathscr{P}_o$**

This section is devoted to the development of the mixing rules for the temperature-independent parameter $\mathscr{P}_o$. These mixing rules may be derived considering the proportionality of $\mathscr{P}_o$ to the intermolecular parameters(or critical properties) according to Kwak and Mansoori (1986). Following the principle of corresponding states a reduced surface tension may be defined as $\sigma_r = \sigma / [T_c^{2/3} (kT_c)^{1/3}]$ and expected to be a universal function of $T_r = T/T_c$ (*i.e.* the reduced temperature). Thus, it can be shown that,

$$\mathscr{P}_{o,ii} \propto T_c^{13/12} / P_c^{5/6} \tag{7}$$

It may be assumed that the temperature-independent parameter, $\mathscr{P}_{o,m}$, for the mixture will follow the same proportionality with respect to the critical temperature and pressure of the mixture,

$$\mathscr{P}_{o,m} \propto \cdot T_{c,m}^{13/12} / P_{c,m}^{5/6} \tag{8}$$

where $T_{c,m}$ and $P_{c,m}$ are the pseudo-critical temperature and pressure for the mixture, respectively. Introducing the van der Waals mixing rules for $T_{c,m}$ and $P_{c,m}$, Eqs. 6a and 6c, into Eq. 8 the following expression is obtained,

$$\mathscr{P}_{o,m} \propto \left\{ \sum_i \sum_j (x_i x_j T_{c_{ij}}^2 / P_{c_{ij}}) / \sum_i \sum_j (x_i x_j T_{c_{ij}} / P_{c_{ij}}) \right\}^{13/12} / \left\{ \sum_i \sum_j (x_i x_j T_{c_{ij}}^2 / P_{c_{ij}}) / [\sum_i \sum_j (x_i x_j T_{c_{ij}} / P_{c_{ij}})]^2 \right\}^{5/6}$$

$$(9)$$

this equation may be further reduced to the following expression,

$$\mathcal{P}_{o,m} \propto \left[\sum_i \sum_j x_i x_j (T_{c_{ij}}{}^2 / P_{c_{ij}})\right]^{1/4} \left[\sum_i \sum_j x_i x_j (T_{c_{ij}} / P_{c_{ij}})\right]^{7/12} \qquad (10)$$

the term $(T_{c_{ij}}{}^2 / P_{c_{ij}})$ maybe further simplified to obtain,

$$T_{c_{ij}}{}^2 / P_{c_{ij}} = (T_{c_{ij}}{}^{11/12} / P_{c_{ij}}{}^{1/6}) \, (T_{c_{ij}}{}^{13/12} / P_{c_{ij}}{}^{5/6}) \qquad (11)$$

notice that the term within the second bracket on the right-hand side of Eq. 11 is nothing but $\mathcal{P}_{o,ij}$ given by Eq. 7.   Thus, Eq. 11 may be rewritten as follows,

$$T_{c_{ij}}{}^2 / P_{c_{ij}} = (T_{c_{ij}}{}^{11/12} / P_{c_{ij}}{}^{1/6}) \, \mathcal{P}_{0,ij} \qquad (12)$$

Eq. 12 can further be rearranged as follows,

$$T_{c_{ij}}{}^2 / P_{c_{ij}} = (P_{c_{ij}} / T_{c_{ij}})^{7/3} \, \mathcal{P}_{0,ij}^4 \qquad (13)$$

By introducing Eq. 13 back into Eq. 10 the following expression for $\mathcal{P}_{o,m}$ is obtained,

$$\mathcal{P}_{o,m} = \cdot\left[\sum_i \sum_j x_i x_j (P_{c_{ij}} / T_{c_{ij}})^{7/3} \, \mathcal{P}_{0,ij}^4\right]^{1/4} \left[\sum_i \sum_j x_i x_j (T_{c_{ij}} / P_{c_{ij}})\right]^{7/12} \qquad (14)$$

Eq. 14 represents the general van der Waals mixing rule proposed in this work for calculation of the temperature-independent parameter, $\mathcal{P}_{o,m}$, for a multicomponent mixture.   The cross-parameters may be calculated using the following expression,

$$\mathcal{P}_{o,ij} = (1 - m_{ij}) \, (\mathcal{P}_{0,ii} \cdot \mathcal{P}_{0,jj})^{1/2} \qquad (15)$$

where $m_{ij}$ is a binary interaction parameter.   The cross-parameters for temperature and pressure in Eq. 14 may be calculated using Eq. 6b and 6d, respectively.

For the case of binary mixtures, the equations to calculate the temperature-independent parameters, $\mathcal{P}^L_{o,m}$ and $\mathcal{P}^V_{o,m}$, for the liquid phase is the following,

$$\mathcal{P}_{0,m}^{L} = \left[ x_{11}^2 \left( P_{c_{11}} / T_{c_{11}} \right)^{7/3} \mathcal{P}_{0,11}^4 + x_{22}^2 \left( P_{c_{\epsilon\epsilon}} / T_{c_{\epsilon\epsilon}} \right)^{7/3} \mathcal{P}_{0,22}^4 + 2x_1 x_2 \left( P_{c_{1\epsilon}} / T_{c_{1\epsilon}} \right)^{7/3} \mathcal{P}_{0,12}^4 \right]^{1/4}$$

$$\times \left[ x_{11}^2 \left( T_{c_{11}} / P_{c_{11}} \right) + x_{22}^2 \left( T_{c_{\epsilon\epsilon}} / P_{c_{\epsilon\epsilon}} \right) + 2x_1 x_2 \left( T_{c_{1\epsilon}} / P_{c_{1\epsilon}} \right) \right]^{7/12} \qquad (16)$$

The temperature-independent parameter $\mathcal{P}_o^y$ for the vapor-phase mixture is calculated by replacing the $x_i$'s in Eq. 16 by the compositions of the vapor phase (*i.e.* $y_i$'s). The cross-parameters are then given by,

$$\mathcal{P}_{o,12} = (1 - m_{12}) \left( \mathcal{P}_{0,1} \cdot \mathcal{P}_{0,2} \right)^{1/2} \qquad (16a)$$

$$T_{c_{12}} = (1 - k_{12}^{RM}) \left( T_{c_1} T_{c_2} \right)^{1/2} \qquad (16b)$$

$$P_{c_{12}} = 8 \, T_{c_{12}} / \left[ \left( T_{c_1} / P_{c_1} \right)^{1/3} + \left( T_{c_2} / P_{c_2} \right)^{1/3} \right]^3 \qquad (16c)$$

Eq. 16 may be used along with Eq. 4 to predict the surface tension of binary mixtures. In the following sections the results for 55 binary mixtures are reported. Table 1 shows the physical properties of the pure fluids involved in these binary mixtures.

The most widely used mixing rules for the temperature-independent parameter, $\mathcal{P}_o$, in Eq. 4 used in the petroleum industry were proposed by Weinaug and Katz (1943) as follows,

$$\mathcal{P}_{o,m} = \sum (x_i \, \mathcal{P}_{o,i}) \qquad (17)$$

Because of the simplicity of this mixing rule, it was decided to investigate the results obtained with Eq. 4 and this simple mixing rule as a preliminary step. Thus, the surface tension of all 55 binary mixtures were predicted using the following procedure (**CASE 1**)

(a) The temperature-independent parameter $\mathcal{P}_o$ for the pure compounds were obtained by fitting Eq. 1 to the experimental surface tension for the pure compounds.
(b) Routine VLE calculations were performed using a computer package developed in our laboratory to obtain the vapor- and liquid-phase compositions as well as the saturation pressure. The results from the VLE calculations were used along with the Riazi-Mansoori equation of state to predict the vapor and liquid densities.
(c) Finally, Eqs. 4 and 17 were used to predict the surface tension for the mixture.

The results obtained for this **CASE 1** are reported in Table 2. Note that the surface tension of all 55 sets of binary-mixture data can be predicted with an overall 2.64 AAD%.

**Surface Tension Prediction for Binary Mixtures**

As mentioned earlier, Eqs. 4, 16, 16a-16c, proposed in this work, represent the necessary expressions to predict the surface tension of binary liquid mixtures. These expressions relate the surface tension of the mixture to the bulk properties of the system such as the vapor- and liquid-phase compositions and vapor- and liquid-phase densities. Therefore, only the temperature-independent parameter $\mathcal{P}_o$ for the pure components is needed. This parameter can be obtained by fitting Eq. 1 to the experimental surface tension data for the pure fluids. It may also be predicted using the corresponding-states expression given in Eq. 3.

In this work, we have predicted the surface tension for 55 binary mixtures and compared the results against the experimental data for these systems. The selected systems represent a wide variety of binary mixtures (symmetric, asymmetric, and slightly asymmetric). The interaction parameter, $m_{ij}$, appearing in Eq. 15 has been treated as an adjustable parameter for all the mixtures investigated.

The surface tension for all the binary mixtures was predicted following the procedure outlined below,

(a) The temperature-independent parameter $\mathcal{P}_o$ for the pure compounds were obtained in two ways: **CASE 2** by fitting Eq. 1 to the experimental surface tension for the pure compounds; and **CASE 3** by predicting them using Eq. 3.

(b) VLE calculations were performed using a computer package developed according to what was discussed above to obtain the vapor- and liquid-phase compositions as well as the saturation pressure. The results from the VLE calculations were used along with the Riazi-Mansoori equation of state to predict the equilibrium vapor and liquid densities.

(c) Finally, Eqs. 4, 16, and 16a-16c were used to predict the surface tension for the mixture.

The results obtained for these two cases are also reported in Table 2. It must be pointed out that the interaction parameter, $m_{ij}$'s in CASE 2 and CASE 3 are different. Therefore, a different value for each case is reported in Table2.

**Results and Discussion for Binary Mixtures**

Table 2 contains the results obtained for **CASES1**, **2**, and **3**. This table also reports temperature, experimental data reference, and number of data points for each binary mixture.

**CASE 2**

The average absolute percent deviation (AAD%) obtained when the binary interaction parameter ($m_{ij}$) was set equal to zero is reported. Note that the surface tension for all the 55 binary mixtures can be predicted within an overall 2.06 AAD%. It may also be noticed that even when $m_{ij}$ is set equal to zero the surface tension for most binary mixtures can be predicted with reasonable accuracy. This represents an improvement over the simple mixing rule (CASE 1)as shown in Table 2. In a different calculation for CASE 2, the binary interaction parameters, $m_{ij}$, were fitted to all binary mixtures. The best value for this parameter is also reported in Table 2. It may be noted that this interaction parameter is small for all the systems investigated. Note also that when this optimized interaction parameter was used the surface tension of all 55 binary mixtures can be predicted within an overall0.50 AAD%. Figures 1 through 9 show the results obtained for CASE 2 when the best value for the binary interaction parameter used in the prediction of surface tension for the binary systems. Note the excellent agreement with the experimental data.

**CASE 3**

The AAD% obtained when the binary interaction parameter ($m_{ij}$) was set equal to zero and the parameters $\mathscr{P}_o$ are predicted using Eq. 3 is reported in Table 2.   Note that the surface tension for all the 55 binary mixtures can be predicted within an overall 3.70 AAD%. It must be pointed out that by setting the binary interaction parameters equal to zero the surface tension is purely predicted since the only adjustable parameter is eliminated.   Note that predictions thus made are in good agreement with the experimental data except in a few cases where Eq. 3 fails to yield good estimates for the parameter $\mathscr{P}_o$.   In another set of calculation for CASE 3 the binary interaction parameter ($m_{ij}$) was fitted to the experimental data. The best values for this parameter are also reported in Table 2.   Note that that when this optimized interaction parameter was used the surface tension of all the 55 binary mixtures can be predicted within an overall 2.10 AAD%.

It should be pointed out that in the absence of surface tension data for pure compounds, Eqs. 3, 4, and 14 (CASE 3) represent an excellent tool to predict the surface tension of mixtures of organic compounds.   Good predictions will be obtained even when the binary interaction parameter, $m_{ij}$, is set equal to zero.

**Conclusions**

Based on the 55 sets of binary-mixture surface tension data analyzed in this work we may conclude that the expressions proposed in this work for surface tension prediction of mixtures are very good.   These predictions are made using only the values of the temperature-independent parameter $\mathscr{P}_o$ for the pure components in the mixture.   This parameter may be obtained from experimental surface tensions of the pure components or estimated using Eq. 3.   It is also worth noting that in these calculations only one adjustable parameter has been used namely the binary interaction parameter, $m_{ij}$, appearing in Eq. 15.

There exist a number of other methods for prediction of the surface tension of liquid mixtures.   None of the other methods are as accurate in predicting mixture surface tension except the ones by Carey *et. al.,* (1980) and Winterfeld *et al.,* (1978).   The latter two methods report very accurate prediction of surface tension of mixtures provided pure fluid surface tension data are available.

It may be concluded that the expressions proposed in this work are general since they have been applied to a variety of binary mixtures and the results obtained are in good agreement with the experimental data.   Application of the present method to multicomponent mixtures is straight forward.   However, little reliable data are available for multicomponent surface tensions for reliable comparisons.

**TABLE 1:** Properties of the pure fluids involved in the 55 binary mixtures studied in this work

| Compound | $T_c$ | $P_c$ | $Z_c$ | $\omega$ | $\mathcal{R}^*$ | $T_b$ |
|---|---|---|---|---|---|---|
| Acetone | 508.1 | 47. | 0.232 | 0.304 | 2.316 | 329.2 |
| Acetonitrile | 545.5 | 48.3 | 0.184 | 0.321 | 1.585 | 354.8 |
| Benzene | 562.2 | 48.9 | 0.271 | 0.212 | 3.748 | 353.2 |
| Carbon Disulfide | 552. | 79. | 0.276 | 0.109 | 3.079 | 319. |
| Carbon Tetrachloride | 556.4 | 49.6 | 0.272 | 0.193 | 3.784 | 349.9 |
| Chloroform | 536.4 | 53.7 | 0.293 | 0.218 | 3.071 | 334.3 |
| Cyclopentane | 511.7 | 45.1 | 0.275 | 0.196 | 3.310 | 322.4 |
| Cyclohexane | 553.5 | 40.7 | 0.273 | 0.212 | 3.966 | 353.8 |
| Cis-Decalin | 702.3 | 32. | 0.245 | 0.286 | 6.281 | 468.9 |
| Trans-Decalin | 687.1 | 31.4 | 0.245 | 0.270 | 6.341 | 460.5 |
| n-Decane | 617.7 | 21.2 | 0.249 | 0.489 | 6.915 | 447.3 |
| Dichloromethane | 510. | 63. | 0.265 | 0.1916 | 2.339 | 313. |
| n-Dodecane | 658.2 | 18.2 | 0.240 | 0.575 | 8.269 | 489.5 |
| Ethyl Acetate | 523.2 | 38.3 | 0.252 | 0.362 | 3.186 | 350.3 |
| Ethyl Ether | 466.7 | 36.4 | 0.262 | 0.281 | 3.219 | 307.6 |
| n-Hexane | 507.5 | 30.1 | 0.264 | 0.299 | 4.281 | 341.9 |
| Iodomethane | 528. | 65.9 | N/A | 0.1493 | 2.788 | 315.7 |
| Iso-Octane | 544. | 25.7 | 0.266 | 0.303 | 5.619 | 372.4 |
| Nitromethane | 588. | 63.1 | 0.208 | 0.310 | 1.787 | 374.3 |
| Phenol | 694.2 | 61.3 | 0.240 | 0.438 | 4.001 | 455. |
| Tetrachloroethylene | 620.2 | 47.6 | 0.250 | 0.254 | 4.341 | 394.4 |
| Toluene | 591.8 | 41. | 0.263 | 0.263 | 4.450 | 383.8 |
| o-Xylene | 630.3 | 37.3 | 0.262 | 0.310 | 5.124 | 417.6 |

The critical properties and acentric factor were obtained from Reid *et al.*, 1986. The parameter $\mathcal{P}_0$ was obtained from experimental surface tension data for pure components; The dimensionless molar refraction $\mathcal{R}^*$ was obtained from data for molar refraction reported by Hall (1986) or calculated using the Lorentz-Lorenz function. For Iodomethane, the acentric factor was estimated by predicting the vapor pressure at Tr = 0.7 using the Antoine equation for vapor pressure prediction. The constants for this equation were obtained from Hall (1986). For Iodomethane there was no experimental critical compressibility factor.

<u>**TABLE 2**</u>    Results obtained for the 55 sets of binary mixtures investigated in this work

| Binary system | CASE 1 AAD% | CASE 2 AAD% $m_{ij}=0$ | fitted $m_{ij}$ | AAD% | CASE 3 AAD% $m_{ij}=0$ | fitted $m_{ij}$ | AAD% | T(K) | Ref. | #data |
|---|---|---|---|---|---|---|---|---|---|---|
| Acetone- Chloroform | 0.42 | 0.48 | 0.0050 | 0.09 | 1.69 | -0.0095 | 0.94 | 291.15 | A | 5 |
| Acetone - Carbon disulfide | 3.06 | 1.42 | 0.0157 | 0.21 | 3.13 | -0.0087 | 2.31 | 288.15 | A | 5 |
| Acetone - Phenol | 3.64 | 3.64 | 0.0305 | 1.72 | 10.24 | 0.0717 | 3.13 | 293.15 | A | 5 |
| Benzene - Acetone | 0.75 | 0.69 | -0.0061 | 0.29 | 5.61 | -0.0326 | 2.24 | 298.15 | B | 9 |
| Benzene - Carbon tetrachloride | 0.22 | 0.22 | 0.0000 | 0.22 | 1.98 | -0.0064 | 1.67 | 298.15 | B | 11 |
| Benzene - Carbon disulfide | 2.77 | 0.31 | 0.0035 | 0.16 | 1.94 | 0.0000 | 1.94 | 291.15 | A | 11 |
| Benzene - Ethyl acetate | 1.00 | 0.69 | 0.0060 | 0.13 | 6.35 | -0.0346 | 2.79 | 298.15 | B | 11 |
| Benzene - Ethyl ether | 2.46 | 2.36 | 0.0202 | 0.44 | 2.21 | 0.0150 | 1.19 | 291.15 | A | 5 |
| Benzene - n-Dodecane | 1.74 | 6.17 | -0.0705 | 0.70 | 7.06 | -0.0659 | 0.96 | 298.15 | C | 5 |
| Benzene - n-Dodecane | 2.37 | 4.79 | -0.0705 | 1.72 | 6.16 | -0.0659 | 1.06 | 303.15 | C | 5 |
| Benzene - n-Dodecane | 1.78 | 4.94 | -0.0705 | 1.30 | 6.01 | -0.0659 | 1.08 | 308.15 | C | 5 |
| Benzene - n-Dodecane | 2.07 | 6.06 | -0.0705 | 1.05 | 6.88 | -0.0659 | 0.71 | 313.15 | C | 5 |
| Benzene - n-Hexane | 4.04 | 3.11 | 0.0275 | 0.97 | 2.16 | 0.0140 | 1.12 | 298.15 | C | 7 |
| Benzene - n-Hexane | 3.70 | 2.86 | 0.0275 | 0.39 | 2.36 | 0.0140 | 1.04 | 303.15 | C | 5 |
| Benzene - n-Hexane | 3.86 | 2.93 | 0.0275 | 0.69 | 2.38 | 0.0140 | 1.01 | 308.15 | C | 5 |
| Benzene - n-Hexane | 3.18 | 2.39 | 0.0275 | 0.63 | 1.80 | 0.0140 | 1.04 | 313.15 | C | 5 |
| Benzene - Nitromethane | 3.47 | 3.69 | 0.0220 | 1.13 | 2.84 | 0.0059 | 2.82 | 298.15 | A | 6 |
| Benzene - o-Xylene | 0.45 | 0.73 | -0.0060 | 0.05 | 2.40 | -0.0143 | 0.73 | 298.15 | D | 10 |
| Benzene - Phenol | 3.50 | 3.52 | 0.0338 | 0.28 | 10.67 | 0.0672 | 4.38 | 308.15 | A | 5 |
| CCl$_4$-Ethyl acetate | 0.69 | 0.50 | 0.0020 | 0.45 | 4.87 | -0.0232 | 2.73 | 298.15 | B | 11 |
| Carbon disulfide – CCl$_4$ | 2.55 | 0.76 | 0.0077 | 0.31 | 4.37 | 0.0296 | 1.45 | 293.15 | E | 10 |
| Carbon disulfide - CCl$_4$ | 2.57 | 0.75 | 0.0077 | 0.25 | 4.39 | 0.0296 | 1.40 | 298.15 | E | 10 |
| Carbon disulfide - CCl$_4$ | 2.76 | 0.94 | 0.0077 | 0.21 | 4.52 | 0.0296 | 1.43 | 303.15 | E | 10 |
| Carbon disulfide - CCl$_4$ | 2.81 | 0.98 | 0.0077 | 0.23 | 4.56 | 0.0296 | 1.49 | 308.15 | E | 10 |
| Carbon disulfide - CCl$_4$ | 2.75 | 0.92 | 0.0077 | 0.30 | 4.47 | 0.0296 | 1.42 | 313.15 | E | 10 |
| Carbon disulfide - CCl$_4$ | 2.91 | 1.08 | 0.0077 | 0.41 | 4.59 | 0.0296 | 1.43 | 318.15 | E | 10 |
| Carbon disulfide - Benzene | 3.56 | 1.29 | 0.0073 | 0.53 | 3.08 | 0.0065 | 2.95 | 298.15 | B | 6 |
| Cyclopentane - CCl$_4$ | 0.29 | 0.31 | 0.0033 | 0.10 | 1.61 | 0.0083 | 0.96 | 298.15 | D | 10 |
| Cyclopentane - Benzene | 2.42 | 2.42 | 0.0182 | 0.19 | 1.31 | 0.0070 | 0.77 | 298.15 | D | 11 |
| Cyclopentane - Toluene | 1.88 | 1.27 | 0.0099 | 0.12 | 0.62 | 0.0035 | 0.44 | 298.15 | D | 12 |
| Cyclopentane-Tetrachloroethylene | 2.01 | 1.84 | 0.0151 | 0.19 | 2.06 | 0.0159 | 0.68 | 298.15 | D | 11 |
| Cyclohexane - Benzene | 2.36 | 1.99 | 0.0169 | 0.14 | 0.82 | 0.0026 | 0.64 | 293.15 | D | 11 |
| Cyclohexane - Benzene | 1.98 | 1.77 | 0.0169 | 0.31 | 1.00 | 0.0026 | 0.98 | 303.15 | D | 11 |
| Cyclohexane - cis-Decalin | 0.67 | 1.12 | -0.0102 | 0.12 | 3.70 | -0.0219 | 1.27 | 298.15 | D | 11 |
| Cyclohexane - trans-Decalin | 0.31 | 1.38 | -0.0113 | 0.11 | 1.76 | -0.0121 | 0.37 | 298.15 | D | 11 |
| Cyclohexane-Tetrachloroethylene | 1.95 | 1.94 | 0.0152 | 0.16 | 1.89 | 0.0134 | 0.80 | 298.15 | D | 11 |
| Cyclohexane - Toluene | 1.93 | 1.86 | 0.0152 | 0.14 | 0.90 | 0.0055 | 0.61 | 298.15 | D | 13 |
| Ethyl ether - Carbon disulfide | 5.27 | 2.49 | 0.0183 | 1.87 | 5.19 | 0.0459 | 4.03 | 291.15 | A | 6 |
| Iodomethane - CCl$_4$ | 2.15 | 1.72 | 0.0118 | 0.99 | 2.90 | 0.0036 | 2.77 | 288.15 | F | 5 |
| Iodomethane - CCl$_4$ | 2.65 | 2.15 | 0.0118 | 0.65 | 2.73 | 0.0036 | 2.60 | 293.15 | F | 8 |
| Iodomethane - CCl$_4$ | 2.57 | 1.93 | 0.0118 | 0.51 | 2.34 | 0.0036 | 2.34 | 298.15 | F | 14 |

<u>**TABLE 2**</u>    (continued)

| Binary system | CASE 1 AAD% | CASE 2 AAD% $m_{ij}=0$ | CASE 2 fitted $m_{ij}$ | CASE 2 AAD% | CASE 3 AAD% $m_{ij}=0$ | CASE 3 fitted $m_{ij}$ | CASE 3 AAD% | T(K) | Ref. | #data |
|---|---|---|---|---|---|---|---|---|---|---|
| Iodomethane - CCl$_4$ | 2.64 | 1.99 | 0.0118 | 0.48 | 2.50 | 0.0036 | 2.47 | 303.15 | F | 11 |
| Iodomethane - CCl$_4$ | 2.37 | 1.80 | 0.0118 | 0.37 | 3.08 | 0.0036 | 3.21 | 308.15 | F | 9 |
| Iso-Octane - Benzene | 3.55 | 1.18 | 0.0096 | 0.58 | 3.00 | 0.0169 | 1.42 | 303.15 | G | 9 |
| Iso-Octane - Cyclohexane | 1.08 | 0.24 | 0.0000 | 0.24 | 2.50 | 0.0101 | 1.50 | 303.15 | G | 5 |
| Iso-Octane - n-Dodecane | 1.16 | 1.59 | -0.0147 | 0.29 | 1.53 | -0.0050 | 1.45 | 303.15 | G | 9 |
| Acetonitrile - Carbon tetrachloride | 1.56 | 1.97 | 0.0160 | 0.82 | 4.57 | -0.0066 | 4.47 | 298.15 | F | 9 |
| Acetonitrile - Carbon tetrachloride | 1.45 | 1.98 | 0.0160 | 0.79 | 4.98 | -0.0066 | 5.02 | 303.15 | F | 8 |
| Acetonitrile - Carbon tetrachloride | 1.75 | 2.10 | 0.0160 | 0.65 | 4.74 | -0.0066 | 4.64 | 308.15 | F | 9 |
| Acetonitrile - Carbon tetrachloride | 1.78 | 2.16 | 0.0160 | 0.54 | 5.59 | -0.0066 | 5.48 | 313.15 | F | 10 |
| Acetonitrile - Carbon tetrachloride | 1.86 | 2.22 | 0.0160 | 0.61 | 5.68 | -0.0066 | 5.58 | 318.15 | F | 9 |
| Dichloromethane-Carbon disulfide | 3.21 | 2.97 | 0.0290 | 0.27 | 4.31 | 0.0304 | 3.04 | 293.15 | H | 8 |
| Dichloromethane-Carbon disulfide | 3.76 | 3.48 | 0.0290 | 0.20 | 4.06 | 0.0304 | 2.80 | 298.15 | H | 9 |
| Dichloromethane-Carbon disulfide | 3.97 | 3.69 | 0.0290 | 0.55 | 4.69 | 0.0304 | 3.15 | 303.15 | H | 10 |
| Dichloromethane-Carbon disulfide | 3.76 | 3.47 | 0.0290 | 0.46 | 4.54 | 0.0304 | 3.32 | 308.15 | H | 10 |
| **Overall AAD%** | **2.64** | **2.06** | | **0.50** | **3.70** | | **2.10** | | | |

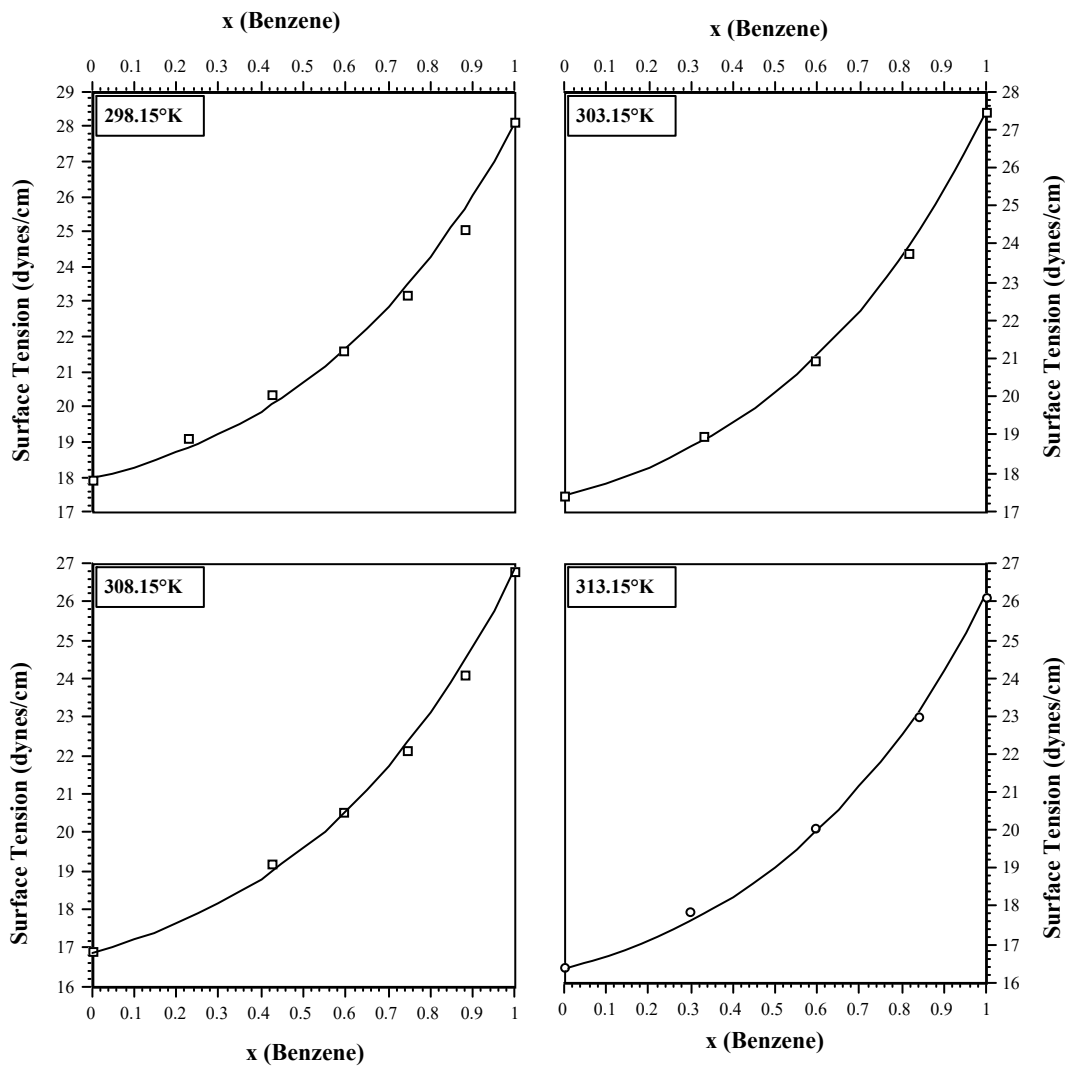

**Figure 1** Surface Tension, σ, versus mole fraction of benzene in the liquid mixture for the system Benzene - n-Hexane. The squares indicate the experimental data (Schmidt *et al.*, 1966). The solid line represents the values calculated with Eqs. 4 and 16.

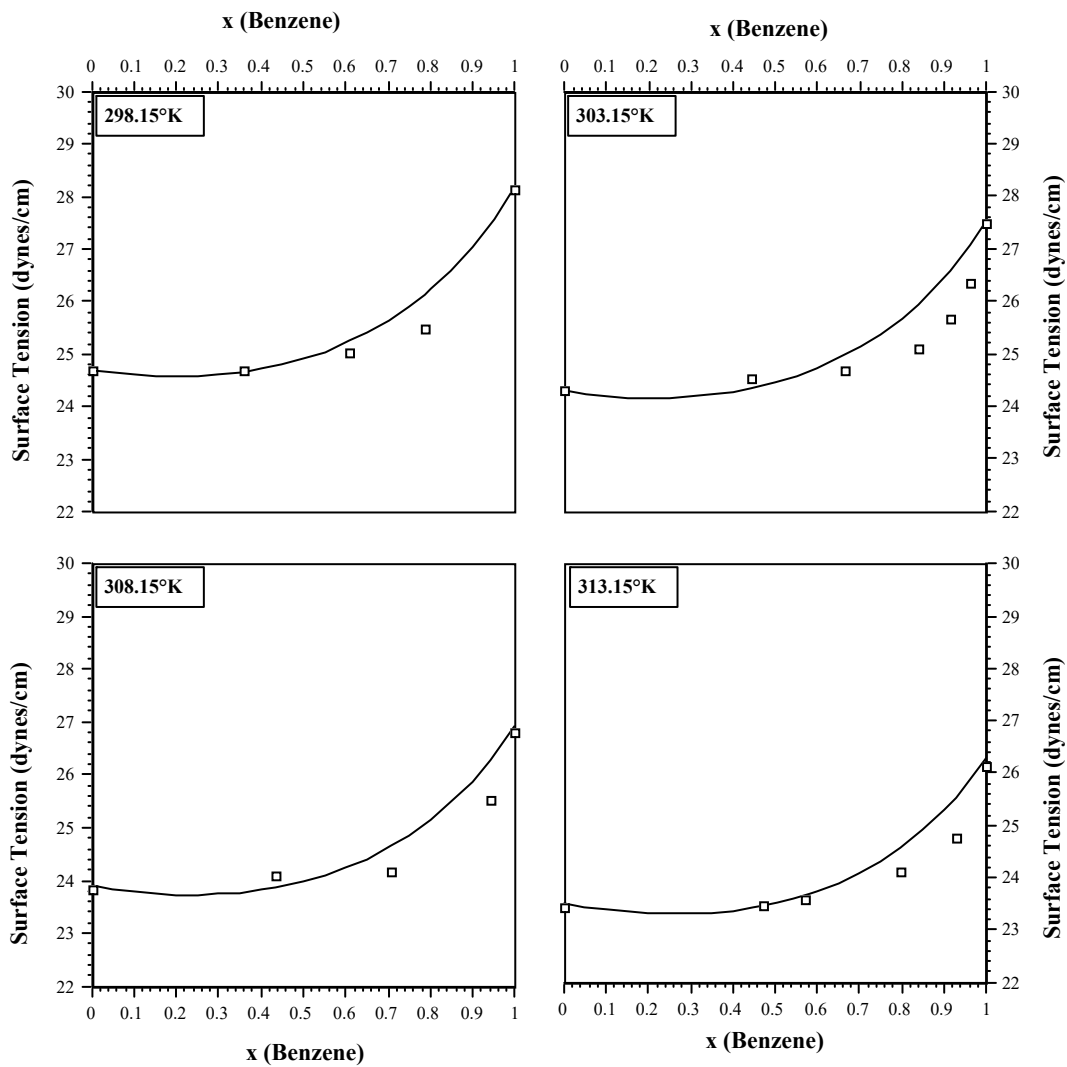

**Figure 2**    Surface Tension, σ, versus mole fraction of benzene in the liquid mixture for the system Benzene - n-Dodecane.   The squares indicate the experimental data (Schmidt *et al.*, 1966).   The solid line represents the values calculated with Eqs. 4 and 16.

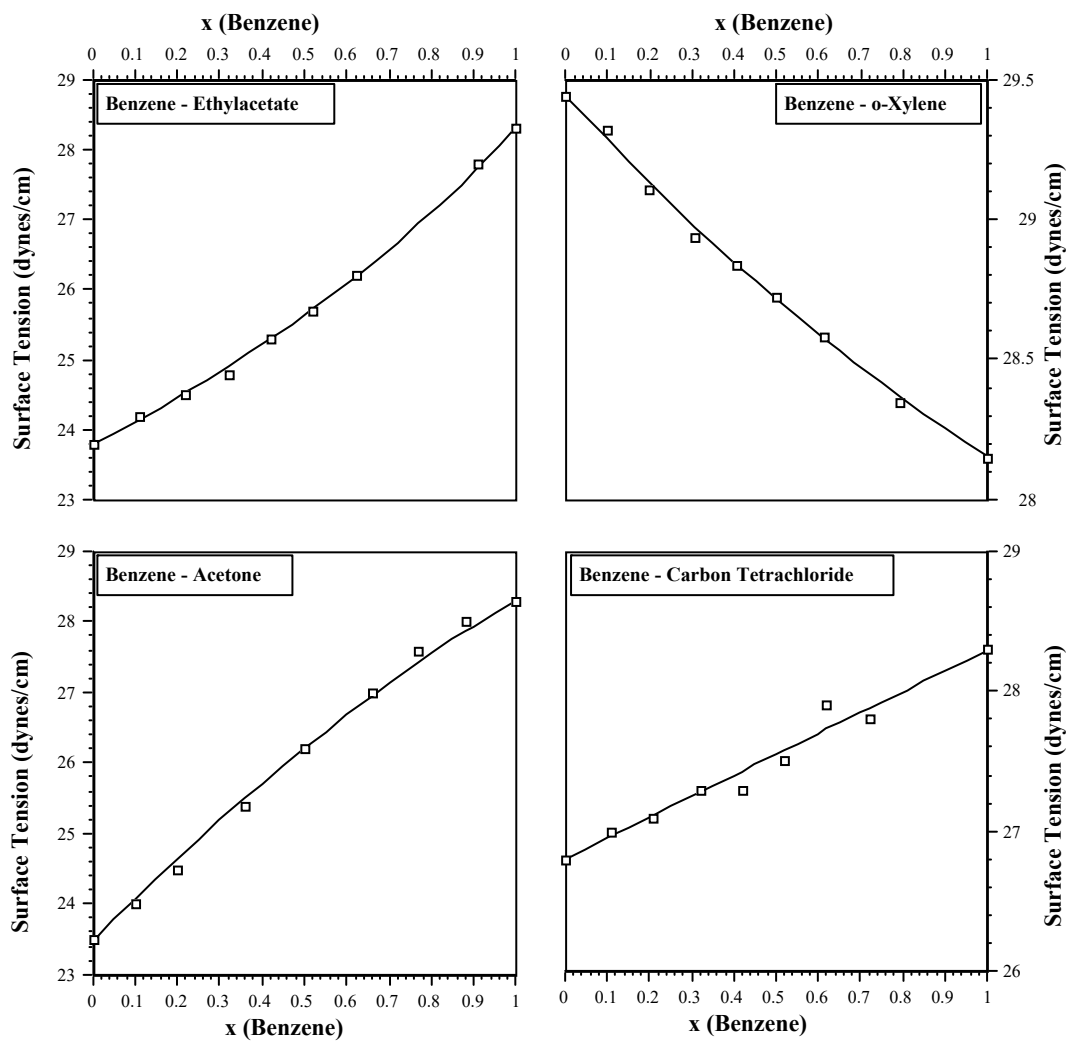

**Figure 3**    Surface Tension, σ, mole fraction of benzene in the liquid mixture for various systems.   The squares indicate the experimental data (Shipp, 1970).  Data for the system benzene - o-xylene was taken from Lam and Benson (1970).   The solid line represents the values calculated with Eqs. 4 and 16.

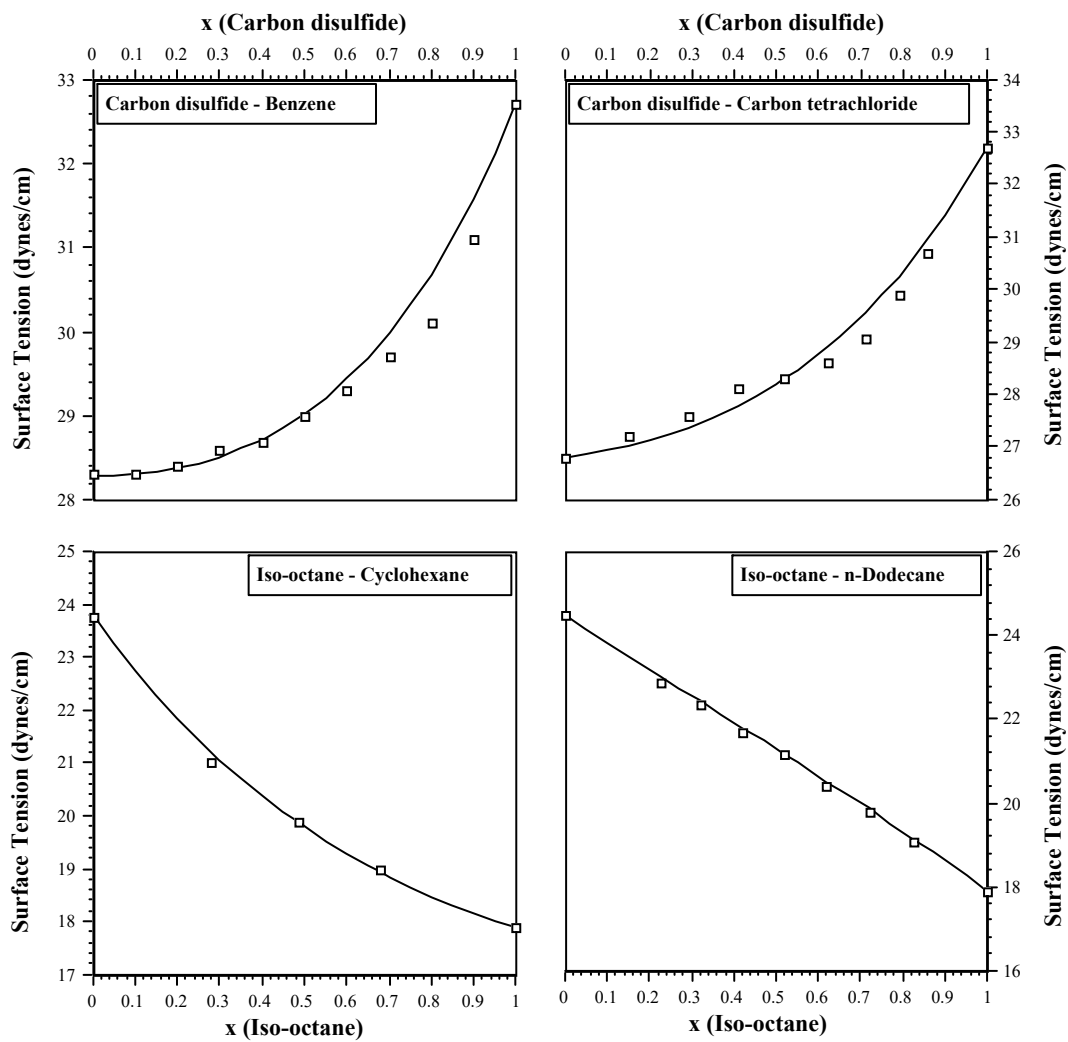

**Figure 4**   Surface Tension, $\sigma$, versus mole fraction for various liquid mixtures.   The squares represent the experimental data.   For the systems with Iso-Octane data are from Evans and Clever (1964); the experimental data for the mixture with carbon disulfide are from Shipp (1970).   The solid line represents the values calculated with Eqs. 4 and 16.

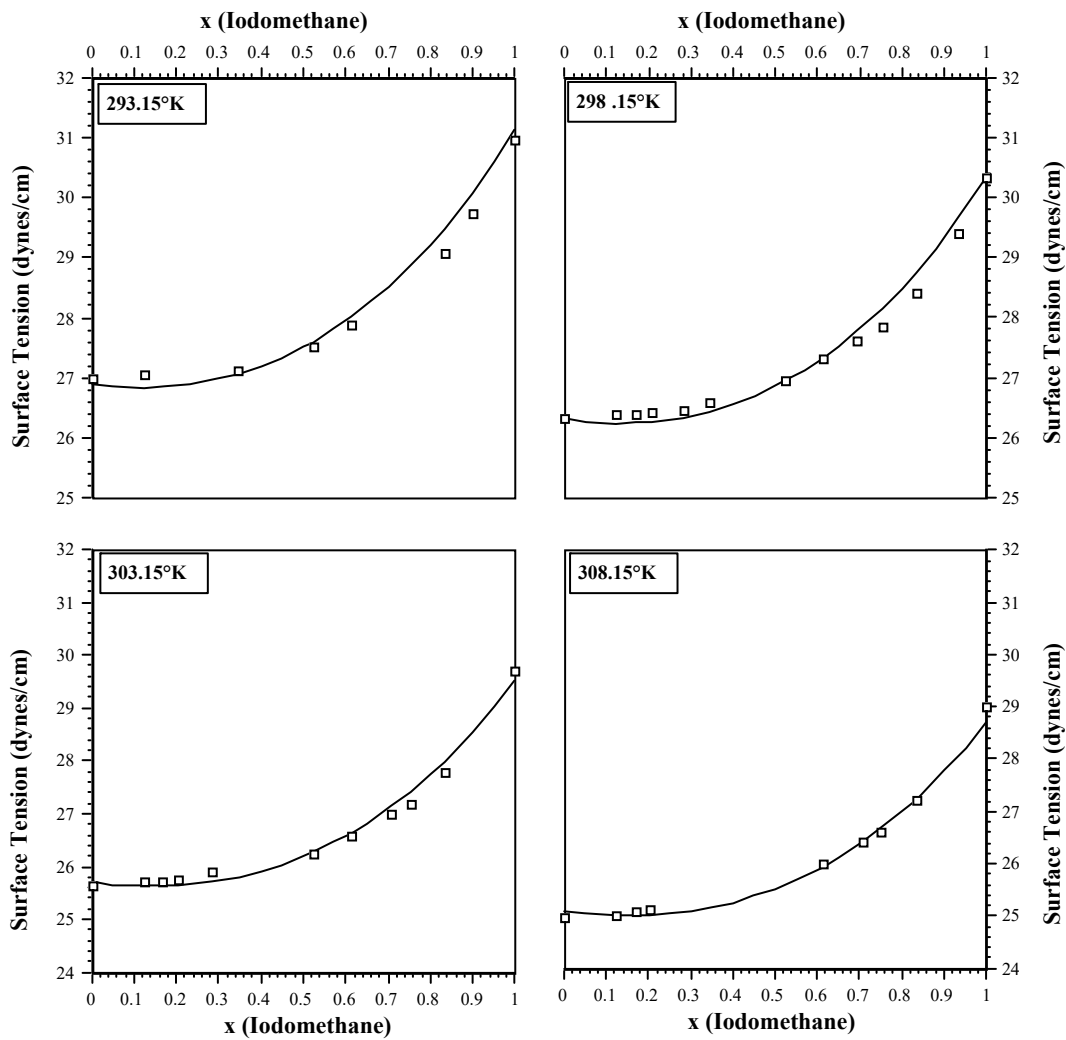

**Figure 5**     Surface Tension, $\sigma$, versus mole fraction of Iodomethane in liquid mixture at various temperatures. The squares indicate the experimental data (Teixeira *et al.*, 1992). The solid line represents the values calculated with Eqs. 4 and 16.

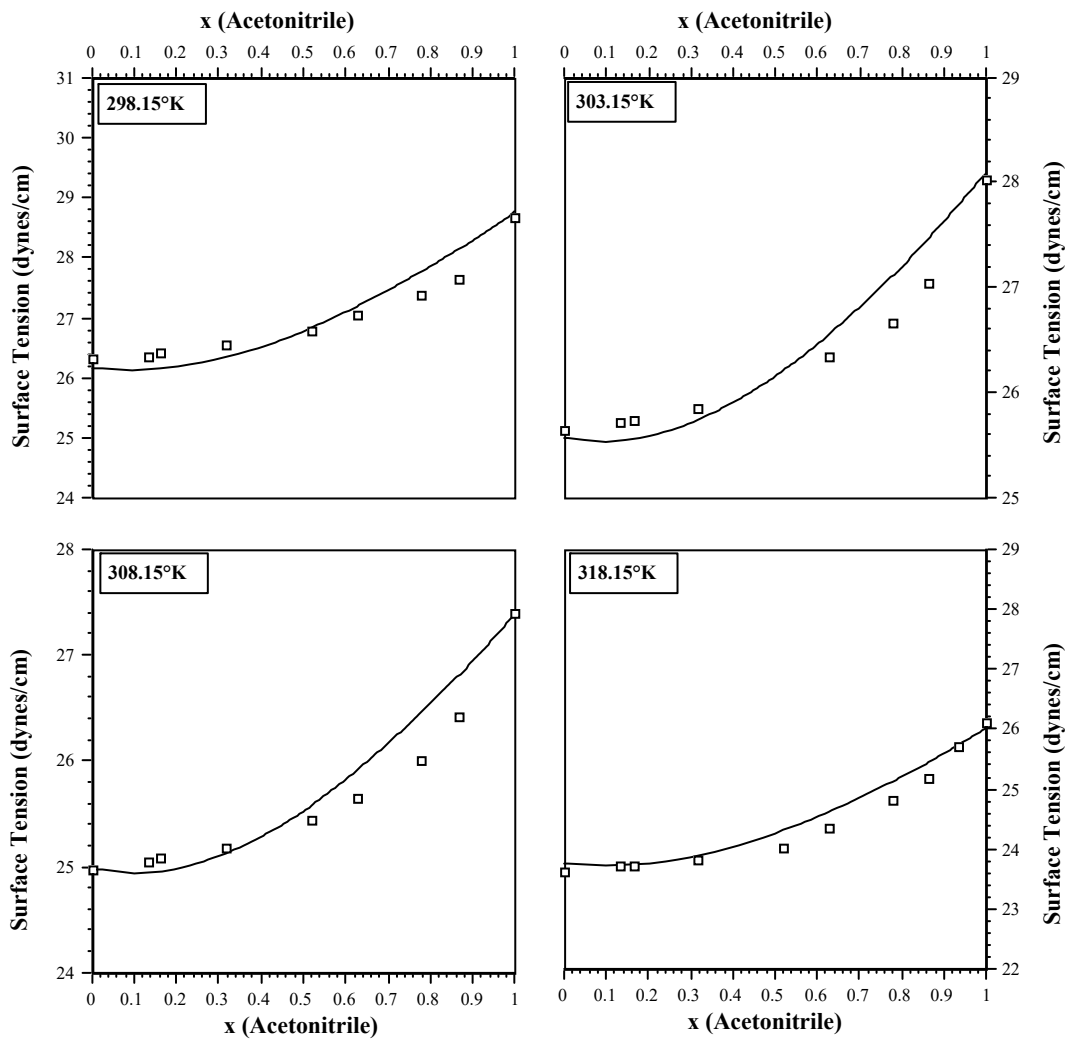

**Figure 6**     Surface Tension, $\sigma$, versus mole fraction of acetonitrile in the liquid mixture at various temperatures. The squares indicate the experimental data (Texeira *et al.*, 1992). The solid line represents the values calculated with Eqs. 4 and 16.

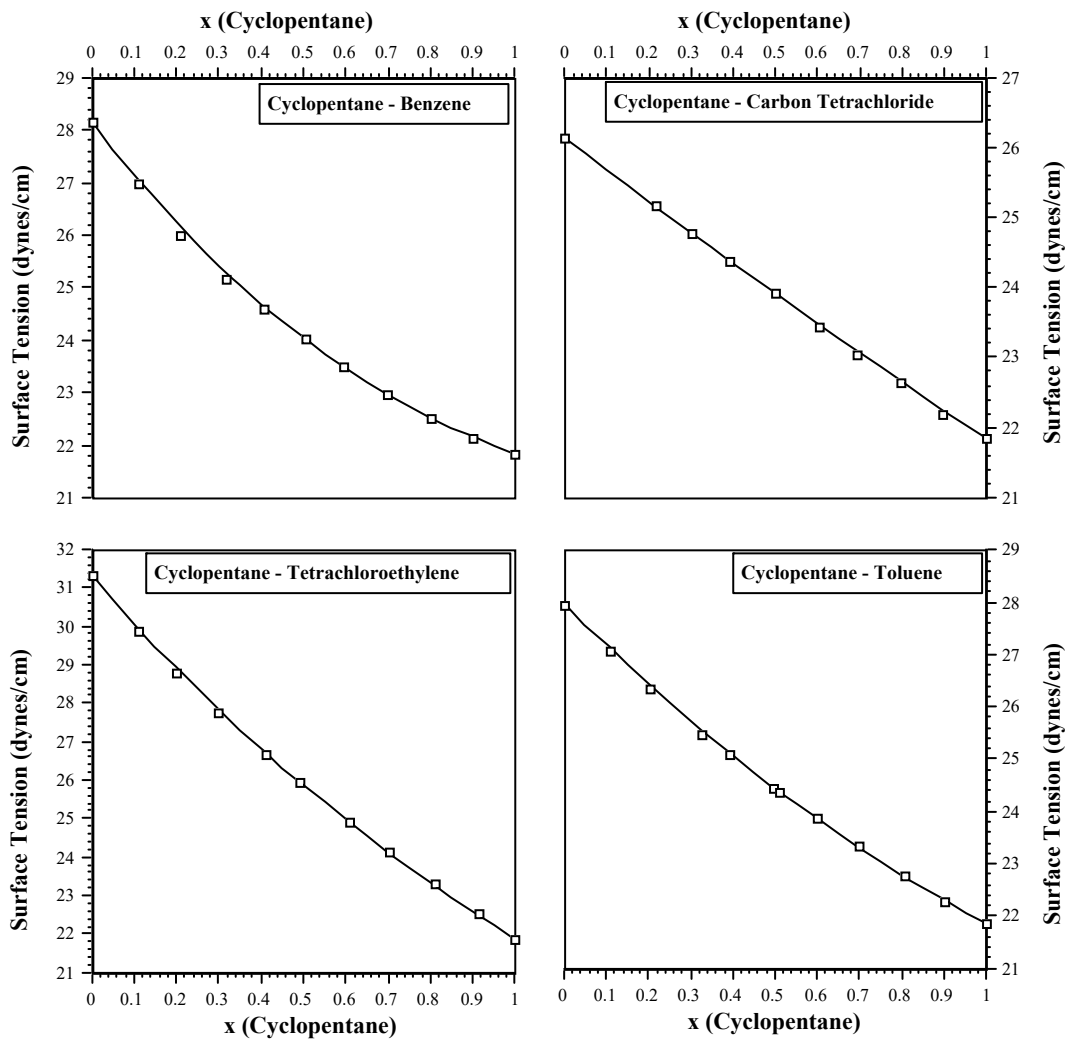

**Figure 7**     Surface Tension, $\sigma$, versus mole fraction of cyclopentane in the liquid mixture for various systems. The squares indicate the experimental data (Lam and Benson, 1970). The solid line represents the values calculated with Eqs. 4 and 16.

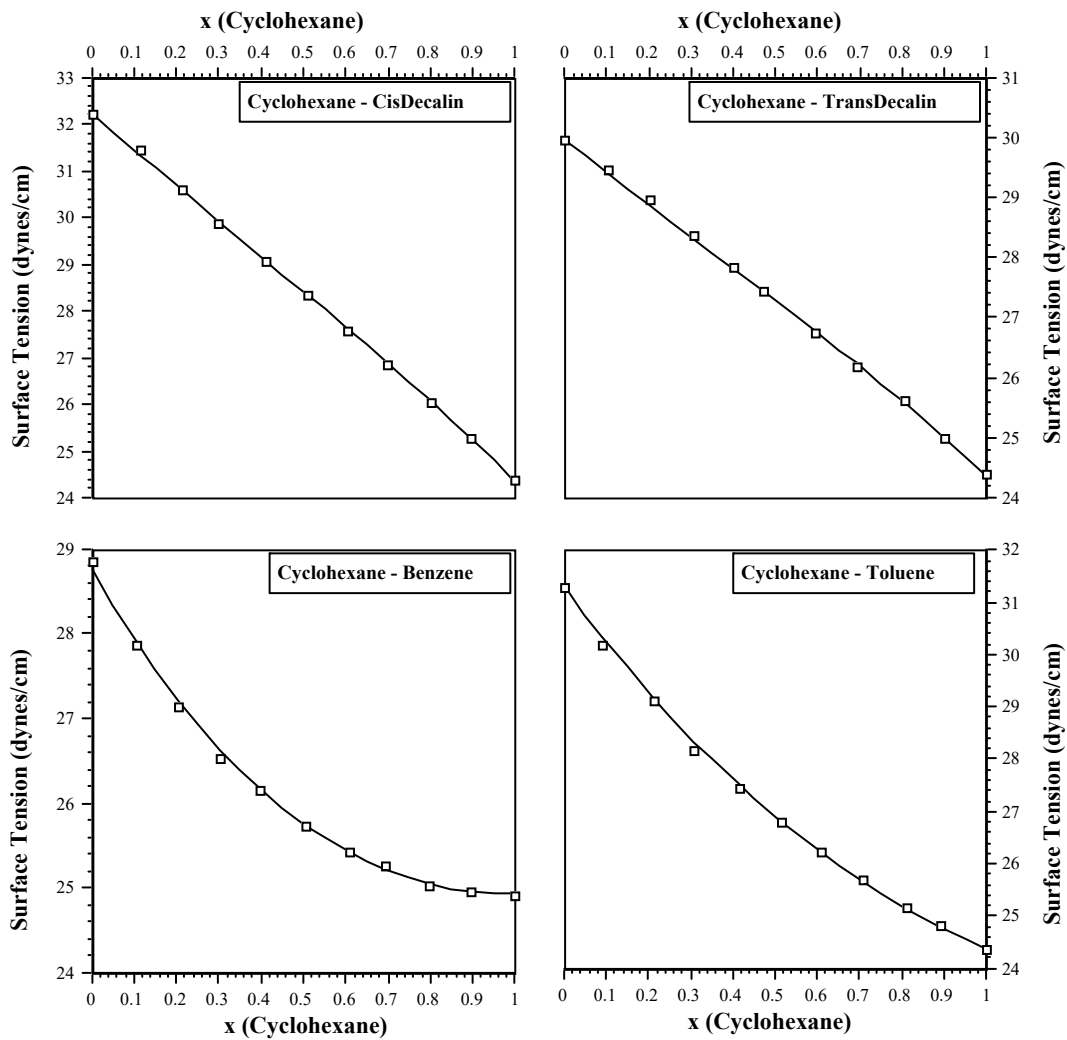

**Figure 8**      Surface Tension, σ, versus mole fraction of cyclohexane in the liquid mixture for various systems. The squares represent the experimental data (Lam and Benson, 1970). The solid line represents the values calculated with Eqs. 4 and 16

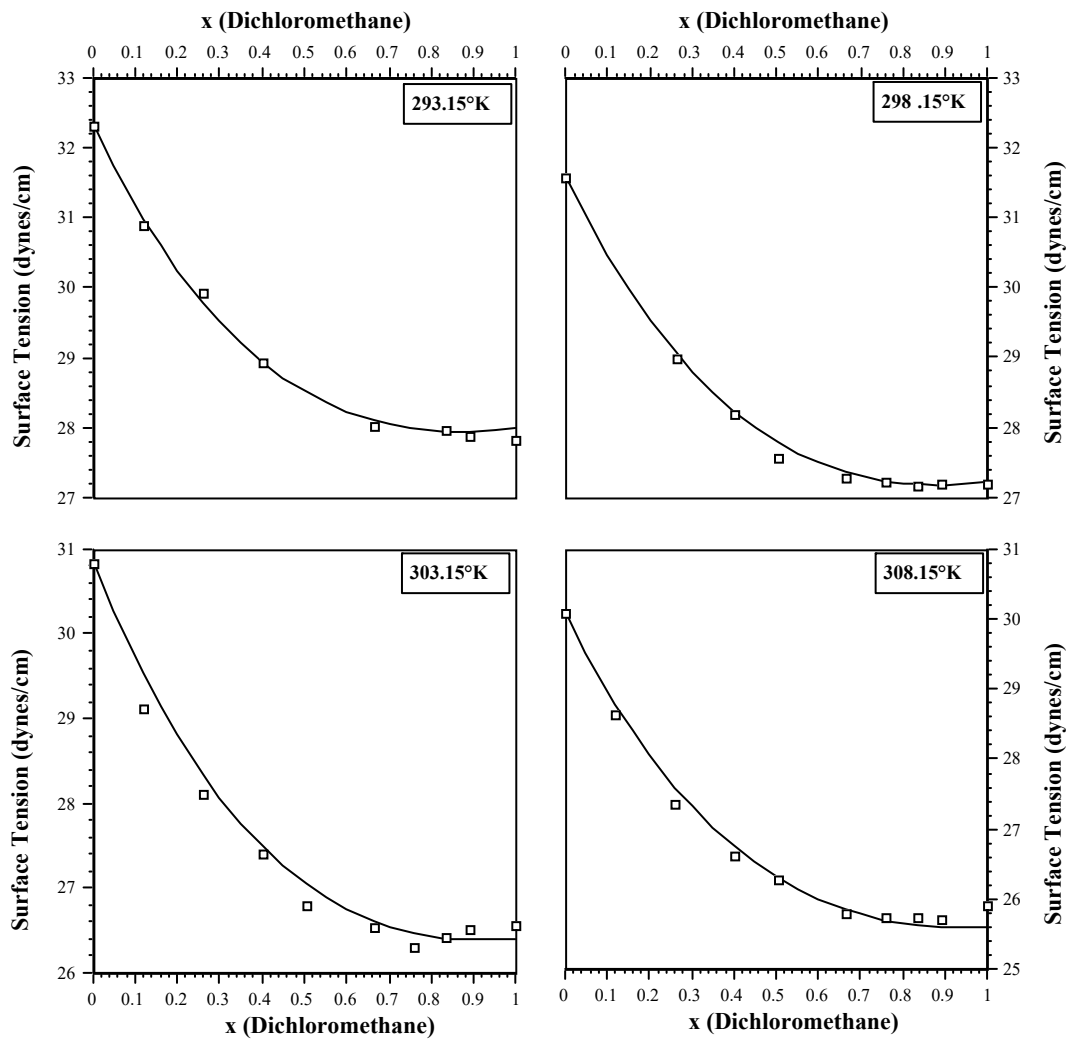

**Figure 9**     Surface Tension, σ, versus mole fraction of dichloromethane in the liquid mixture at various temperatures. The squares indicate the experimental data (Aracil *et al.*, 1989).   The solid line represents the values calculated with Eqs. 4 and 16